\documentclass[reprint,aps,pra,twocolumn,superscriptaddress,floatfix,longbibliography]{revtex4-2}

\usepackage[dvips]{graphicx}
\usepackage{latexsym}
\usepackage{amsmath}
\usepackage{amsthm}
\usepackage{amsfonts}
\usepackage{amssymb}
\usepackage{mathptmx}
\usepackage{subfigure}
\usepackage{mathrsfs}

\usepackage{color}
\definecolor{darkblue}{rgb}{0,0.02,0.45}
\usepackage[colorlinks=true,citecolor=darkblue]{hyperref}

\usepackage[all]{hypcap}
\usepackage{gensymb}

\definecolor{cream}{RGB}{222,217,201}

\begin{document}

\title{Role of alkaline metal in the rare-earth triangular antiferromagnet KYbO$_2$}

\author{Franziska Gru{\ss}ler}
\email{franziska.grussler@physik.uni-augsburg.de}
\affiliation{Experimental Physics VI, Center for Electronic Correlations and Magnetism, University of Augsburg, 86135 Augsburg, Germany}

\author{Mamoun Hemmida}
\affiliation{Experimental Physics V, Center for Electronic Correlations and Magnetism, University of Augsburg, 86135 Augsburg, Germany}

\author{Sebastian Bachus}
%\email{lei.ding.ld@outlook.com}
\affiliation{Experimental Physics VI, Center for Electronic Correlations and Magnetism, University of Augsburg, 86135 Augsburg, Germany}

\author{Yurii Skourski}
\affiliation{High Magnetic Field Laboratory (HLD-EMFL, Helmholtz-Zentrum Dresden-Rossendorf, 01328 Dresden, Germany}

\author{Hans-Albrecht Krug von Nidda}
\affiliation{Experimental Physics V, Center for Electronic Correlations and Magnetism, University of Augsburg, 86135 Augsburg, Germany}

\author{Philipp Gegenwart}
\affiliation{Experimental Physics VI, Center for Electronic Correlations and Magnetism, University of Augsburg, 86135 Augsburg, Germany}

\author{Alexander~A.~Tsirlin}
\email{altsirlin@gmail.com}
\affiliation{Experimental Physics VI, Center for Electronic Correlations and Magnetism, University of Augsburg, 86135 Augsburg, Germany}
\affiliation{Felix Bloch Institute for Solid-State Physics, Leipzig University, 04103 Leipzig, Germany}

\begin{abstract}
We report crystal structure and magnetic behavior of the triangular antiferromagnet KYbO$_2$, the A-site substituted version of the quantum spin liquid candidate NaYbO$_2$. The replacement of Na by K introduces an anisotropic tensile strain with 1.6\% in-plane and 12.1\% out-of-plane lattice expansion. Compared to NaYbO$_2$, both Curie-Weiss temperature and saturation field are reduced by about 20\% as the result of the increased Yb--O--Yb angles, whereas the $g$-tensor of Yb$^{3+}$ becomes isotropic with $g=3.08(3)$. Field-dependent magnetization shows the plateau at $\frac12$ of the saturated value and suggests the formation of the up-up-up-down field-induced order  in KYbO$_2$ in contrast to the more common $\frac13$ plateau with the up-up-down order that has been reported in the isostructural Yb$^{3+}$ selenides. 
\end{abstract}

\maketitle

\section{Introduction} 
Since Anderson proposed the formation of a quantum spin liquid (QSL) in triangular antiferromagnets~\cite{Anderson1973}, the scientific community struggled to find a real-life material to investigate this intriguing state of matter. QSL boasts high spin entanglement combined with strong quantum fluctuations and absence of magnetic order down to the lowest temperatures. Furthermore, fractionalized excitations of the QSL may have far-reaching implications for quantum technology~\cite{Wen2019,Broholm2020,Chamorro2021}.

In 2015, the research on triangular QSLs was enlivened by the discovery of YbMgGaO$_4$~\cite{Li2015PRL}. The absence of magnetic order down to at least 48\,mK~\cite{Li2016,Ding2020}, the $C_v\propto T^{2/3}$ power-law behavior of the specific heat~\cite{Li2015,Paddison2017}, and the observation of an excitation continuum in the inelastic neutron scattering spectra~\cite{Shen2016,Li2017b,Li2019} marked YbMgGaO$_4$ as a promising example of the U(1) QSL. However, effects of structural randomness strongly influence this putative QSL behavior because the disorder in the non-magnetic Mg/Ga layer creates a random charge environment of the Yb$^{3+}$ ions, thus leading to a distribution of local magnetic moments and exchange interactions~\cite{Li2017,Zhu_2017,Li_2020}.

Despite these potential complications, YbMgGaO$_4$ sparked the interest in the rare-earth-based triangular antiferromagnets, especially those with Yb$^{3+}$ as the pseudospin-$\frac12$ magnetic ion. Triangular antiferromagnets AYbX$_2$ (A = Li, Na, K, Rb, Cs and X = S, Se, O) have become the focus of research over the past years~\cite{Baenitz2018,Ding2019,Ranjith2019,Bordelon2019,Sarkar2019,Schmidt2021}, since most of them are believed to be free from any structural disorder. With more and more members of the AYbX$_2$ family being prepared and investigated~\cite{Ranjith2019,Zhang2021,Dai2021,Zhang2021b,Zhu2021,Xing2019,Xing2020,Xing2021,Scheie2021,Haussler2022}, one seeks to understand in what manner the single-ion magnetism of Yb$^{3+}$ and the interactions in the triangular Yb$^{3+}$ layers are affected by the surrounding ions. The YbMgGaO$_4$ case has shown that the surrounding non-magnetic atoms are not simple observers. They can have crucial influence on the low-temperature behavior and formation of a QSL~\cite{Li_2020}.

\begin{figure} [b]
\centering
\includegraphics[width=1\linewidth]{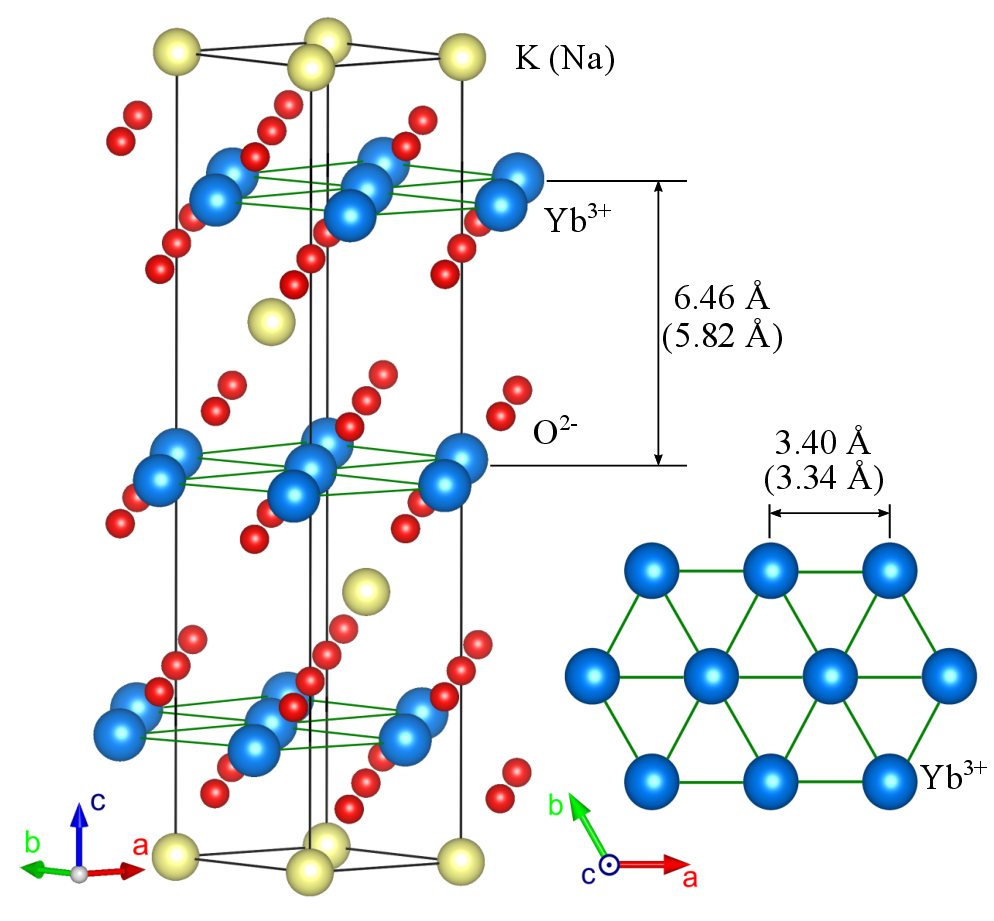}
\caption{Schematic crystal structures of KYbO$_2$ and NaYbO$_2$. The values for the inter- and intralayer Yb--Yb distances for KYbO$_2$ are determined from the crystal structure refinement at 10\,K, see below. The values for NaYbO$_2$~\cite{Ding2019} are given in brackets.}\label{fig:1}
\end{figure}

Here, we address this question by a detailed comparison between NaYbO$_2$, reported previously as the disorder-free QSL candidate~\cite{Ding2019,Bordelon2019,Ranjith2019}, and isostructural KYbO$_2$~\cite{Dong2008} that has not been on radar of magnetism research. The Li compound with the same stoichiometry is also known, but it features a different crystal structure and, consequently, the three-dimensional diamond spin lattice~\cite{Bordelon2021}.

\renewcommand{\arraystretch}{1.5}

\begin{table*}[t]
\centering
\caption{Comparison of NaYbO$_2$ and KYbO$_2$. The lattice parameters, Yb-O and (Na/K)-Yb distances are given in \r A, while the bridging angles Yb-O-Yb are expressed in degrees. The powder-averaged $g$-value is determined from ESR. The van Vleck contribution to the susceptibility, $\chi_\mathrm{vv}$, is determined from the linear part of the high-field magnetization and given in units of emu/mol. The characteristic temperature $\theta$ (in K) is obtained from the Curie-Weiss fits after subtracting $\chi_{\rm vv}$. The values for NaYbO$_2$ are taken from Ref.~\cite{Ding2019} and from Ref.~\cite{Ranjith2019} ($g_{\rm ESR}$).}
\begin{tabular*}{\textwidth}{@{\extracolsep{\fill}}c c c c c c c c c}\hline\hline
{} &{$a$}&{$c$} &{Yb-O dist.}&{$\measuredangle$Yb-O-Yb}&{(Na/K)-Yb dist.}&{$g_\mathrm{ESR}$}&{$\chi_\mathrm{vv}$}&{$\theta$}\\
\hline
NaYbO$_2$ &{3.34481(4)}&{16.4585(2)} &{2.25537(3)}&{95.723(1)°}&{3.35466(3)}&{2.86(7)}&{0.00564(6)}&{$-6.4(3)$\,K} \\

KYbO$_2$ &{3.39731(4)}&{18.453(3)} &{2.27232(3)}&{96.756(1)°}&{3.64786(5)}&{3.08(3)}&{0.00399(1)}&{$-5.4(2)$\,K} \\
\hline
\hline
\end{tabular*}
\label{abc}
\end{table*}

\renewcommand{\arraystretch}{1}

In NaYbO$_2$, Yb$^{3+}$ features a Kramers doublet ground state separated from the first excited doublet by $\Delta=34.8\,\mathrm{meV}\approx 400\,$K~\cite{Ding2019,Bordelon2019}. Below 50\,K, an effective spin-$\frac12$ description holds, albeit with the anisotropic $g$-tensor~\cite{Ranjith2019}. Magnetic interactions between the effective spins are on the order of 5\,K, yet specific heat and muon spin relaxation measurements confirm the absence of magnetic long-range order and the presence of spin dynamics down to at least 70\,mK in zero field~\cite{Ding2019,Bordelon2019}. Long-range magnetic order is induced in the magnetic fields above 3\,T~\cite{Ranjith2019,Bordelon2019}. In the following, we track how these features evolve upon replacing sodium in NaYbO$_2$ with potassium, and thus reveal the role of alkaline metals in the AYbX$_2$ family of triangular antiferromagnets.

\begin{figure*}[t]
\centering
\includegraphics[width=0.98\linewidth]{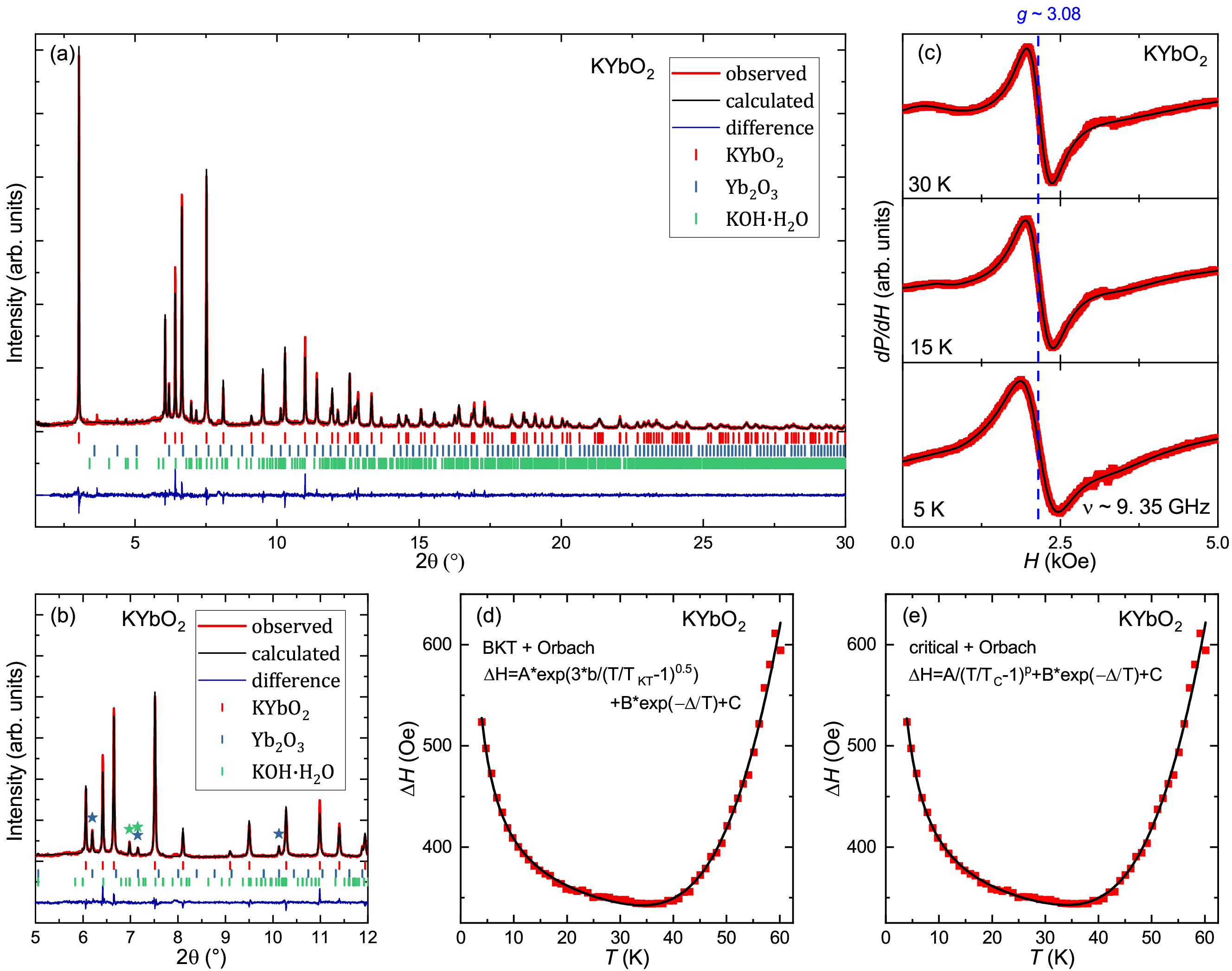}
\caption{(a) Structure refinement using the synchrotron diffraction pattern at 10\,K. (b)  Zoom in to the angular range between 5 and 12$^\circ$ where the reflections from the impurity phases are marked with stars.  (c) ESR spectra of KYbO$_2$ taken at the X-band frequency for selected temperatures. The solid line indicates the fit with the field derivative of a symmetric Lorentz line. {(d,e) Fits of the ESR linewidth with an Orbach process at high temperatures and two possible models at low temperatures (see text for details).}}\label{fig:2}
\end{figure*}

\section{Methods}
\textit{Sample preparation.} Polycrystalline KYbO$_2$ was synthesized by a solid-state reaction using KO$_2$ and Yb$_2$O$_3$ as starting materials. The reagents were weighed and ground in an agate mortar in an argon-filled glovebox to prevent a reaction of KO$_2$ with air moisture. An excess of 75\% KO$_2$ was used similar to the approach in Ref.~\cite{Dong2008}. The powdered reagents were filled into a platinum crucible, placed in a horizontal furnace, heated to 650 °C, held at this temperature for 16\,h, and subsequently cooled down to room temperature. The reaction was performed using an argon flow of 20\,sccm. The finished product was stored under argon atmosphere since it rapidly decomposes in contact with air moisture. For a direct comparison with KYbO2, 
the polycrystalline sample of NaYbO$_2$ was prepared similar to the previous study~\cite{Ding2019}.

\textit{Crystal structure.} Synchrotron X-ray diffraction measurements were performed to determine the crystal structure (Fig.~\ref{fig:1}). Diffraction data were collected at the MSPD beamline~\cite{Fauth2013} (ALBA, Spain) using the multianalyzer setup and the wavelength of $\lambda = 0.32525$\,\r A at the constant temperature of 10\,K. To reduce preferred-orientation effects, the powder sample was placed into a thin-wall glass capillary and spun during the measurement.
\renewcommand{\arraystretch}{1.5}%
\begin{table}[htp]
\centering
\caption{Atomic positions and atomic displacement parameters $U_{\rm iso}$ determined from the structure refinement against the 10\,K synchrotron data. The results for NaYbO$_2$ are from Ref.~\cite{Ding2019}.}
\begin{tabular*}{0.4\textwidth}{@{\extracolsep{\fill}}c |c c l |c}
\hline
\hline
\multicolumn{5}{l} {NaYbO$_2$} \\
\hline
{atoms} &{$x/a$}&{$y/b$} &{$z/c$}&{$U_\mathrm{iso}$(\r A$^2$)}\\
\hline
O &{0}&{0} &{0.2375(1)}&{0.0007(4)} \\

Yb &{0}&{0} &{0.5}&{0.00015(3)} \\
Na &{0}&{0} &{0}&{0.0032(3)} \\
\hline
\hline
\multicolumn{5}{l} {KYbO$_2$} \\
\hline
{atoms} &{$x/a$}&{$y/b$} &{$z/c$}&{$U_\mathrm{iso}$(\r A$^2$)}\\
\hline
O &{0}&{0} &{0.2288(3)}&{0.005(1)} \\

Yb &{0}&{0} &{0.5}&{0.0015(2)} \\
K &{0}&{0} &{0}&{0.0046(5)} \\
\hline
\hline
\end{tabular*}
\label{ADPs}
\end{table}
\renewcommand{\arraystretch}{1}

\textit{Magnetic characterization.} Temperature-dependent magnetic susceptibility was measured using the SQUID magnetometer from Quantum Design (MPMS 3) from 2 to 300\,K at 1\,T (ZFC) and between 0.4 and 2\,K  in various magnetic fields between 1 and 5\,T utilizing the $^3$He cooling stage. 
Field-dependent magnetization measurements up to  7\,T were performed at various temperatures between 0.4\,K and 1\,K using the MPMS with the $^3$He cooling stage. The Quantum Design PPMS equipped with the vibrating sample magnetometer option provided an extension of the magnetic field up to 14\,T at 2\,K. The field-dependent magnetization up to 40\,T was measured at 0.54\,K using a triply compensated extraction magnetometer within a 50\,T mid-length-pulse magnet at the High Magnetic Field Laboratory Dresden.

\begin{figure*}
\centering
\includegraphics[width=0.98\linewidth]{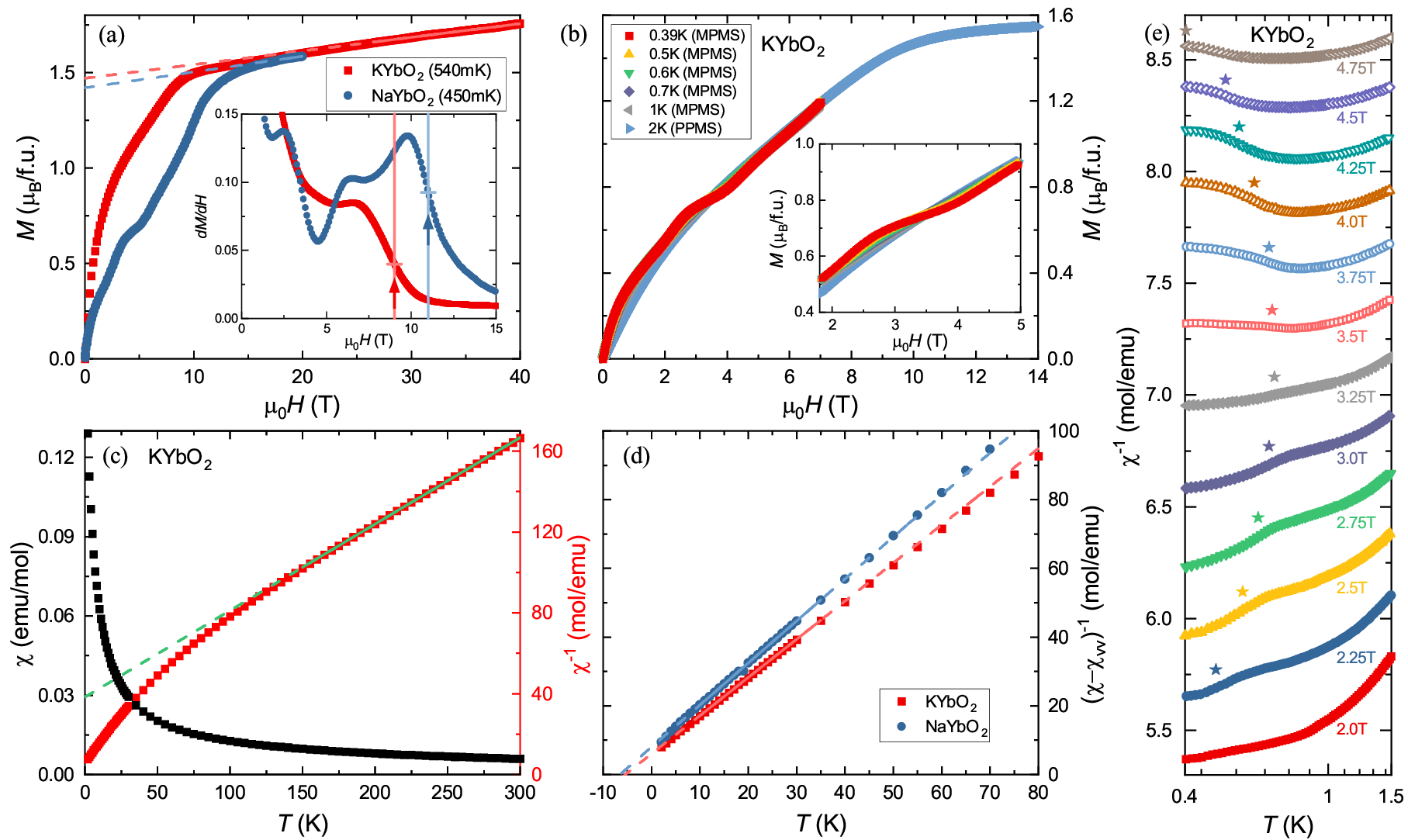}
\caption{(a) Magnetization of KYbO$_2$ and NaYbO$_2$   measured at 0.54\,K up to 40\,T and 0.45\,K up to 20\,T, respectively. The linear fits are used to determine the Van Vleck contribution. The magnetization data of NaYbO$_2$ was adapted from \cite{Ding2019} by scaling the high-field measurement to our $^3$He MPMS measurements (see Fig.~\ref{fig:7}). The inset shows the field derivative of the magnetization with the arrows indicating the respective saturation fields. (b) Magnetization of KYbO$_2$ measured with the MPMS and PPMS (blue triangles, 2\,K) at various temperatures. At low temperatures, a plateau is clearly observed between 2 and 5\,T at about half of the saturation magnetization. (c) Magnetic susceptibility and inverse magnetic susceptibility for KYbO$_2$ measured in the applied field of 1\,T. The green line represents the high-temperature Curie-Weiss fit. (d) Low-temperature Curie-Weiss fit for NaYbO$_2$ and KYbO$_2$ after the subtraction of the van Vleck term $\chi_\mathrm{vV}$. (e) Low-temperature susceptibility of KYbO$_2$ measured at various magnetic fields. The transition temperatures are indicated by stars. }  \label{fig:3}
\end{figure*}

\textit{Specific heat measurements.} The specific heat of KYbO$_2$ as well as NaYbO$_2$ was measured by thermal relaxation method between 0.5 and 30\,K using the PPMS from Quantum Design with the $^3$He insert at various magnetic fields. Zero-field measurements were extended up to 300\,K using the standard $^4$He setup. In the case of NaYbO$_2$, a dilution refrigerator has been used to extend the temperature range down to 70\,mK. The specific heat in fields up to 8\,T has been determined by a thermal relaxation method. Further details can be found in Ref.~\cite{Ding2019}, where selected low-field data ($B \le 1.25$\,T) for NaYbO$_2$ have already been reported.

\textit{ESR measurements.}
The electron spin resonance (ESR) measurements were performed in a continuous wave spectrometer (Bruker ELEXSYS E500) at X-band frequency ($\nu \approx 9.35 $\,GHz) in the temperature region $4\leq T\leq 300$\,K using a continuous He gas-flow cryostat (Oxford Instruments).
ESR detects the power $P$ absorbed by the sample from the transverse magnetic microwave field as a function of the static magnetic field $H$. The signal results from magnetic dipole transitions between the Zeeman levels of the electron spins. The signal-to-noise ratio of the spectra is improved by recording the derivative $dP/dH$ using a lock-in technique with field modulation. The measurement was done on a polycrystalline sample fixed in a quartz tube with paraffin.

\section{Results}

\subsection{Crystal structure}

Our synchrotron XRD data confirm the $R\bar 3m$ space group and the high crystallinity of KYbO$_2$. Two minor impurity phases also found in the sample, Yb$_2$O$_3$ and KOH $\cdot$ H$_2$O (Fig.~\ref{fig:2}(a) and (c)), are decomposition products, as the capillary was shortly exposed to air upon the insertion into the cryostat. Crystal structure refinement reveals low atomic displacement parameters (ADPs), most notably, the low ADP of Yb, in contrast to YbMgGaO$_4$ where Yb ADP remains sizable, about 0.01\,\r A$^2$ at 100\,K~\cite{Li2015PRL}, and signals local displacements caused by the random distribution of Mg$^{2+}$ and Ga$^{3+}$ in the structure. On the other hand, the KYbO$_2$ structure should be fully ordered, similar to NaYbO$_2$.

In Tables~\ref{abc} and \ref{ADPs}, we compare the structural parameters of NaYbO$_2$ and KYbO$_2$. The replacement of Na by K leads to a highly anisotropic lattice expansion with the 1.6\% increase in $a$ and 12.1\% increase in $c$. This anisotropy can be traced back to the rigid nature of the [YbO$_2$] layers that undergo only a weak expansion demonstrated by the 0.8\% increase in the Yb--O distances. This expansion is insufficient to account for the 1.6\% increase in the Yb--Yb distances, so the Yb--O--Yb angles increase too. They, in turn, affect the local environment of Yb$^{3+}$, because the Yb--O--Yb and O--Yb--O angles are equal by symmetry. 

Overall, we find that negative pressure introduced by the K substitution causes both expansion (increase in the Yb--O distances) and flattening (increase in the Yb--O--Yb angles) of the triangular [YbO$_2$] layers. In contrast, hydrostatic pressure effect reported for YbMgGaO$_4$~\cite{Majumder2020} leads to a uniform compression wherein the Yb--O distances shrink while the Yb--O--Yb angles remain almost constant. The tensile strain introduced by the chemical substitution is thus highly anisotropic.

\subsection{$g$-values}

Local magnetism of the Yb$^{3+}$ ions is revealed by the ESR measurements. After subtracting the cavity background, the Yb-ESR lines are well described by the field derivative of a single Lorentzian [see Fig.~\ref{fig:2}(b)]. Distortions between 2.8 and 3.5\,kOe are due to the fact that the background  could not be perfectly subtracted, because of slightly different quality factor with and without the sample. The broad shoulder at low fields was taken into account by a second Lorentzian derivative, but it is ascribed to the impurity phase.  On increasing temperature, the ESR signal becomes very broad due to an Orbach process that includes population of excited crystal-electric-field (CEF) levels of Yb$^{3+}$~\cite{Sichelschmidt2020}. 

Interestingly, only one symmetric ESR line is observed in KYbO$_2$. This stands in contrast to the previous reports on other AYbX$_2$ compounds. The $g$-tensors of both NaYbS$_2$ and NaYbSe$_2$ show a strong easy-plane anisotropy, with $g_{\|}\leq 1.0$ and $g_{\perp}\simeq 3.2$~\cite{Sichelschmidt2019,Ranjith2019,Haussler2022}. A similar easy-plane anisotropy is observed in NaYbO$_2$ where $g_{\|}=1.75(3)$ and $g_{\perp}=3.28(8)$ were reported on a polycrystalline sample~\cite{Ranjith2019_NaYbO2}. On the other hand, our data suggest an almost isotropic $g\simeq 3.08(3)$ in KYbO$_2$. This change parallels the structural evolution of the [YbO$_2$] layers that undergo an anisotropic expansion. Consequently, the composition of the ground-state Kramers doublet of Yb$^{3+}$ should change, and its $g$-values are modified accordingly.

\begin{figure*}
\centering
\includegraphics[width=\linewidth]{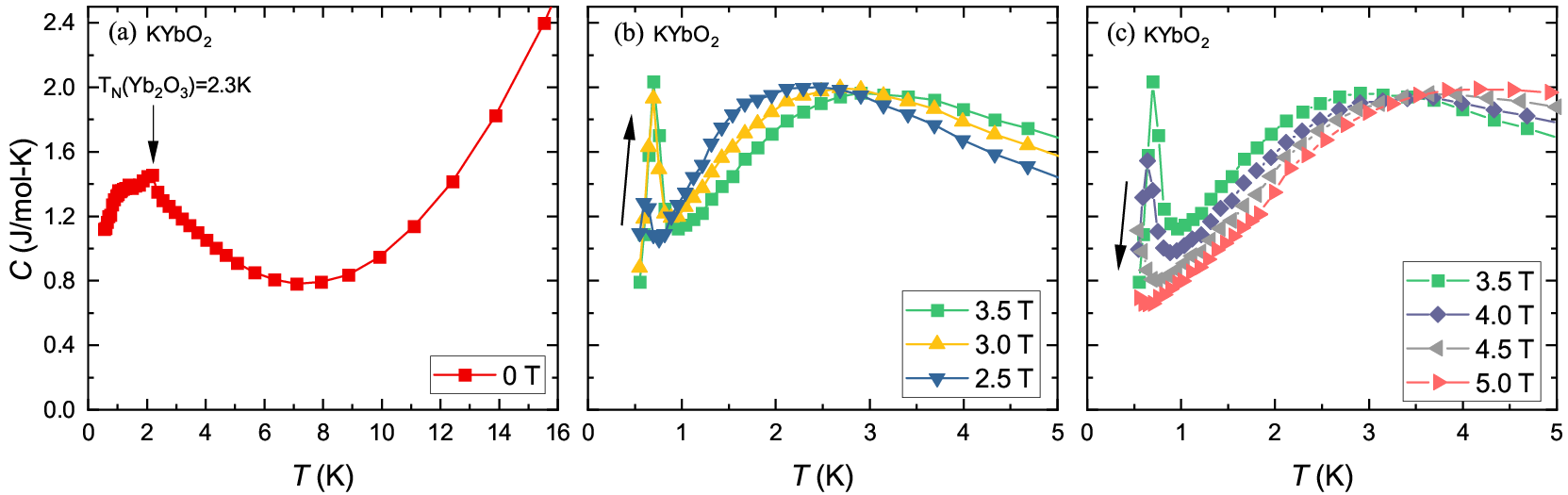}
\caption{(a) Specific heat of KYbO$_2$ measured in zero field. The 2.3\,K anomaly corresponds to the Yb$_2$O$_3$ impurity phase. (b) Magnetic order appears for magnetic fields larger than 2.5\,T. The peak becomes more pronounced with increasing the field up to 3.5\,T. (c) For fields larger than 3.5\,T the peak becomes smaller and shifts to lower temperatures.}
  \label{fig:4}
\end{figure*}

Temperature dependence of the ESR linewidth is shown in Fig.~\ref{fig:2}(d) and (e). The broadening at high temperatures is ascribed to the aforementioned Orbach process, $\Delta H \propto \exp (-\Delta /T)$. The resulting energy gap $\Delta\simeq 350$\,K between the ground-state Kramers doublet and first excited CEF level is quite close to that determined for NaYbO$_2$ ($\Delta=350\,$K \cite{Sichelschmidt2020} and $\Delta=320\,$K \cite{Ranjith2019_NaYbO2}).

{ At low temperatures, two possible models are considered. One interpretation relies on the classical critical behavior in the vicinity of a phase transition, $\Delta H \propto (T/T_\mathrm{c}-1)^{-p}$ with the critical temperature $T_\mathrm{c}$ [Fig.~\ref{fig:2}(e)]. The other interpretation [Fig.~\ref{fig:2}(d)] assumes the proximity to a Berezinskii-Kosterlitz-Thouless (BKT) transition with $\Delta H \propto \exp[3b/((T/T_\mathrm{KT})-1)^{0.5}]$, the Kosterlitz-Thouless (KT) temperature $T_\mathrm{KT}$, and $b=\pi/2$~\cite{Heinrich2003}. The BKT interpretation was previously applied to the ACrO$_2$ triangular antiferromagnets (A = H, Li, Na)~\cite{Hemmida2009}, although later studies challenged this scenario~\cite{Somesh2021}. While there are no direct indications for a significant XY anisotropy in KYbO$_2$, we used both models to demonstrate: i) equally good fits that make it hard to conclude on the applicability of the BKT scenario from the ESR data alone; ii) similarity of the critical temperatures, $T_\mathrm{KT}=0.14$\,K and $T_\mathrm{c}=0.15$\,K, extracted from both fits. We also note that the critical exponent $p=0.77$ is similar to the $T^{-0.75}$ power-law behavior established in sibling compounds~\cite{Sichelschmidt2020,Sichelschmidt2019}.}

%The fits are visibly indistinguishable and have nearly identical values for $\chi^2$ and $R^2$. From the fit at higher temperatures following the Orbach process  The low-temperature ESR linewidth of other rare-earth based triangular antiferromagnets \cite{Sichelschmidt2020,Sichelschmidt2019} was only investigated in the pathway of the classical critical behavior, where a $T^{-0.75}$ power-law behavior was determined. This is in agreement with the exponent $p=0.77$ determined here. The KT temperature $T_\mathrm{KT}=0.14$\,K and the critical temperature $T_\mathrm{c}=0.15\,K$ determined from the respective fits call for investigations at temperatures below the available temperature range of our measurements. Note that the critical scenario needs 6 fit parameters, i.e. 3 for the critical term, two for the Orbach process and a residual line width, while the BKT scenario needs only 5 parameters, i.e. 2 of the BKT term, when fixing $b=1.5$ and the exponent $p=0.5$ and the same number of parameters for Orbach process and residual width. For the BKT scenario it is even possible to fix the residual line width at zero, which results in a reasonable fit with a somewhat lower Kosterlitz-Thouless temperature but still comparably low $\chi^2$ value. In contrast, fixing the residual line width at zero in case of the critical scenario yields an unrealistically high $T_\mathrm{c}$ above 2K. Thus, the BKT scenario is favorable to describe the low-temperature behavior of KYbO$_2$ in terms of a topological phase transition. 

\begin{figure*}
\centering
\includegraphics[width=\linewidth]{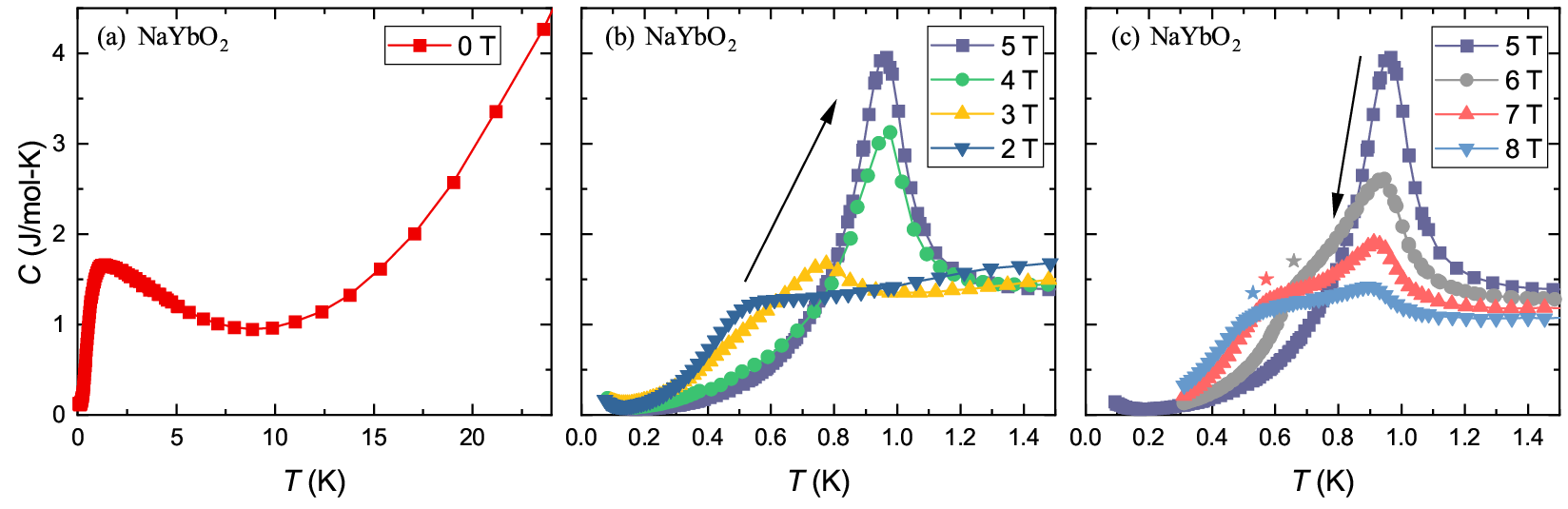}
\caption{(a) Measurements of the specific heat of NaYbO$_2$ in zero field. (b) Magnetic order is first observed in a magnetic field of 3\,T. The transition is shifted to higher temperatures with increasing magnetic field up to 5\,T. (c) Further increasing the magnetic field leads to a decrease in the transition temperature. The stars mark the lower-temperature shoulder that appears above 6\,T.}
  \label{fig:6}
\end{figure*}

\subsection{Thermodynamic properties}

Inverse magnetic susceptibility of KYbO$_2$ shows linear behavior at high temperatures and gradually changes slope below 150\,K [see Fig.~\ref{fig:3}(c)]. The Curie-Weiss fit of the linear part returns the Curie-Weiss temperature $\theta= -89.6\,$K and the effective moment of $\mu_\mathrm{eff}=4.32\mu_\mathrm{B}$, which is similar to the free-electron value of 4.54$\mu_\mathrm{B}$ for Yb$^{3+}$. 

At low temperatures, the ground-state Kramers doublet of Yb$^{3+}$ is responsible for the pseudospin-$\frac12$ behavior, whereas the effect of higher-lying CEF levels is taken into account by the temperature-independent van Vleck term $\chi_\mathrm{vv}$~\cite{Li_2020}. We determine its value from field-dependent magnetization that shows the linear increase above the saturation field [Fig.~\ref{fig:3}(a)]. In KYbO$_2$, the saturation is observed around 9\,T, whereas the linear part at high magnetic fields ($27-40$\,T) yields $\chi_\mathrm{vv}=0.00715$\,$\mu_\mathrm{B}$/T$=0.00399$\,emu/mol. The saturation magnetization of 1.47\,$\mu_\mathrm{B}$/f.u. can be deduced after correcting $M(H)$ for the van Vleck contribution. This value is in a good agreement with the saturation magnetization $M_\mathrm{sat,ESR}=g\,S=1.54$\,$\mu_\mathrm{B}$/f.u. calculated using the $g$-value from ESR and $S=\frac12$.

In order to gauge magnetic interactions between the ground-state Kramers doublets, we subtract $\chi_{\rm vv}$ from the experimental susceptibility and find the linear regime in $(\chi-\chi_{\rm vv})^{-1}$ between 10 and 50\,K [Fig.~\ref{fig:3}(d)]. The corresponding effective moment of 2.68$\mu_\mathrm{B}$ perfectly matches the expected value of $g\sqrt{S(S+1)}=2.67$\,$\mu_B$ with $g=3.08$ and $S=\frac12$, while the negative Curie-Weiss temperature of $\theta_{CW}=-5.4$\,K indicates antiferromagnetic interactions.

It is instructive to compare $(\chi-\chi_{\rm vv})^{-1}$ between KYbO$_2$ and NaYbO$_2$ [Fig.~\ref{fig:3}(d)]. The K compound shows a lower slope because of the higher effective moment, which is determined by $g=3.08$, to be compared with the powder-averaged $g_{\rm av}=2.86$ in NaYbO$_2$. On the other hand, the absolute value of the Curie-Weiss temperature is reduced upon replacing Na with K. A similar trend can be seen in the saturation field that decreases from 11\,T in NaYbO$_2$~\cite{Ranjith2019_NaYbO2,Ding2019} to about 9\,T in KYbO$_2$.

We now turn to the low-temperature behavior and discuss the plateau feature, which is observed in $M(H)$ below 0.7\,K. This slightly tilted plateau is clearly seen in the MPMS data measured in static fields below 0.7\,K [Fig.~\ref{fig:3}(b)]. The absence of the plateau in the pulsed-field data [Fig.~\ref{fig:3}(a)] may be due to the slight sample heating during the pulse. Magnetization plateaus have been observed in many of the AYbX$_2$ compounds when ordered spin states are induced by the applied field. In AYbSe$_2$ (A = Na, K, Rb, Cs), measurements on single crystals showed that the plateau appears for in-plane magnetic fields only and corresponds to $\frac13$ of the saturation magnetization~\cite{Ranjith2019,Xing2021}. Neutron scattering experiments in the applied field confirmed up-up-down (\textit{uud}) nature of the magnetic order~\cite{Scheie2022,Xie2022}. The same \textit{uud} order was observed by neutrons in NaYbO$_2$~\cite{Bordelon2019}, although magnetization at the plateau reaches $\frac12$ of the saturation value~\cite{Ding2019} and exceeds expected magnetization of the \textit{uud} state. This discrepancy might be caused by the anisotropic nature of the Yb$^{3+}$ moments. Interestingly, magnetization of KYbO$_2$ at the plateau, about 0.7\,$\mu_B$/f.u., also reaches $\frac12$ of the saturation value. Given the isotropic nature of the Yb$^{3+}$ moments, this result speaks against the formation of the \textit{uud} phase. An up-up-up-down (\textit{uuud}) field-induced order would be a more likely candidate. 

The field-induced magnetic order also manifests itself as a kink in the low-temperature susceptibility shown in Fig.~\ref{fig:3}(e). For fields below 3.5\,T a downward turn in $\chi^{-1}(T)$ is observed, while for fields larger than 3.5\,T {inverse susceptibility shows an upward curvature}. The transition temperature is { determined via the respective maximum or minimum in the first derivative of $\chi^{-1}(T)$ and increases} from 0.49\,K in an applied field of 2.25\,T to 0.71\,K at 3.25\,T. Above 3.25\,T the transition temperature decreases and disappears below the measurement range above 4.75\,T.

Field-induced magnetic order is further monitored by specific heat measurements performed with the $^3$He insert down to 0.5\,K (see Fig.~\ref{fig:4}). No magnetic order is observed in zero field, similar to NaYbO$_2$. Specific heat reveals a broad maximum indicative of short-range magnetic order, as well as a small anomaly around 2.3\,K due to the magnetic ordering transition in the Yb$_2$O$_3$ impurity phase~\cite{Moon1967}.
  
Magnetic field systematically shifts the specific heat maximum toward higher temperatures, the same can be observed for NaYbO$_2$ (see Fig.~\ref{fig:6}). Above 2\,T, field-induced long-range magnetic order in KYbO$_2$ is revealed by a $\lambda$-type anomaly. The ordering peak shifts to higher temperatures on increasing the field up to 3\,T, then shifts to lower temperatures on further increasing the field up to 4.5\,T. At 5\,T, the onset of the ordering peak is still visible but the peak itself lies below our measurement range. A qualitatively similar behavior is observed in NaYbO$_2$ (Fig.~\ref{fig:6}).

\begin{figure*}[htbp]
\includegraphics[width=0.85\linewidth]{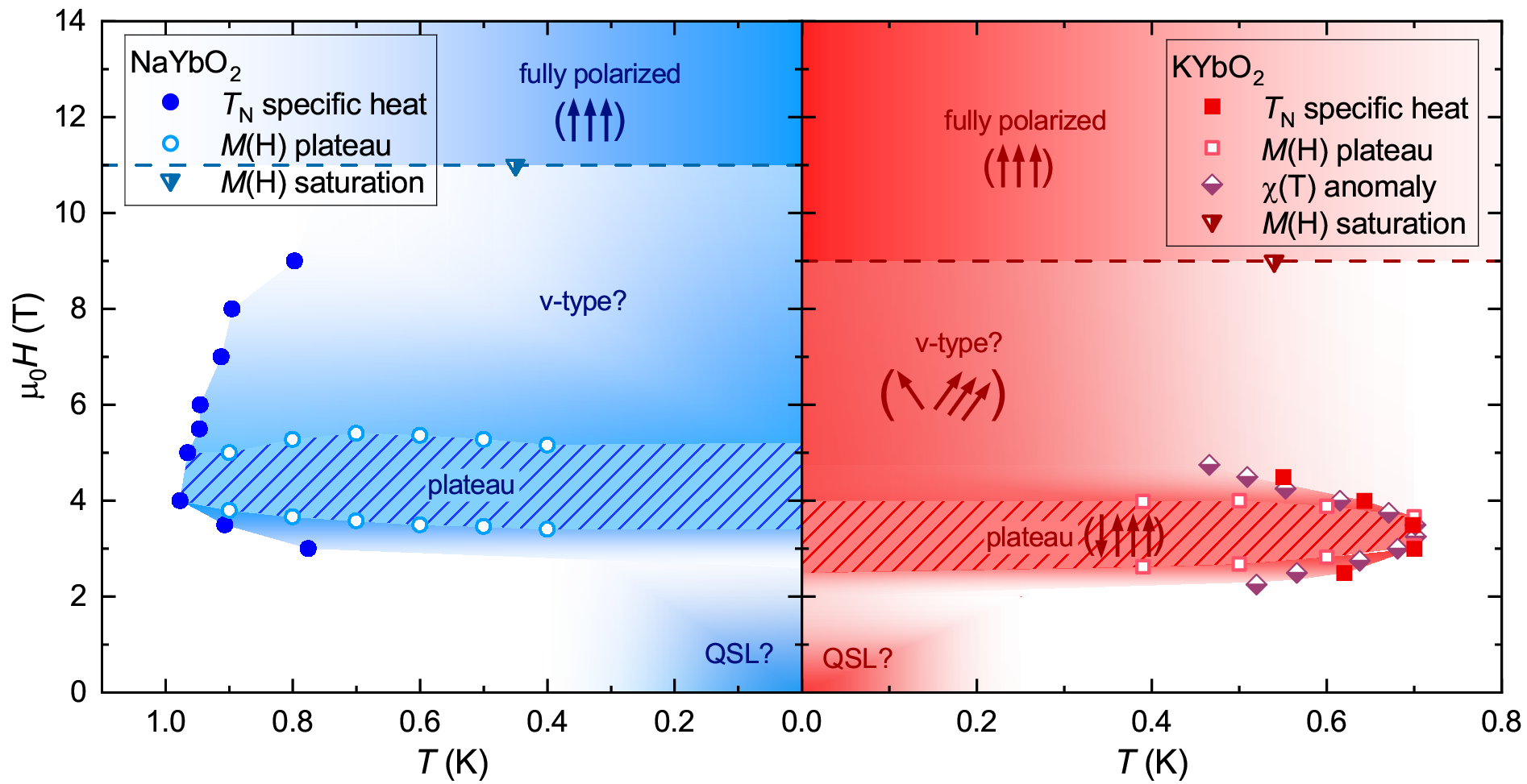}
\caption{Comparison of the field-temperature phase diagrams for NaYbO$_2$ (blue, left) and KYbO$_2$ (red, right) derived from the specific heat (filled symbols), field-dependent magnetization (open symbols) measurements and temperature-dependent magnetic susceptibility (partially filled symbols) measurements.}  \label{fig:5}
\end{figure*}

By tracing the $\lambda$-type anomaly in different magnetic fields, the start and end of the magnetization plateau ({ determined from the maxima in the second derivative of $M(H)$}), and the kink in the magnetic susceptibility, we construct field-temperature phase diagrams for both KYbO$_2$ and NaYbO$_2$ (Fig.~\ref{fig:5}). They show that the stability region of the plateau phase in the K compound is merely scaled down with respect to the Na one, in line with the weaker magnetic interactions in the former. 

Above 5\,T, NaYbO$_2$ displays a clear splitting between the plateau end measured by $M(H)$ and the magnetic ordering transition measured by $C_p(T)$. This splitting indicates that another ordered phase should appear above the plateau region. It may be a noncollinear V-type phase that has been pinpointed in Co-based triangular antiferromagnets~\cite{Liu2019} and also proposed theoretically~\cite{Ye_2017O,Ye_2017}. The KYbO$_2$ data show no clear signatures of such a phase, although the temperature range of our measurements was probably insufficient for detecting it. Experiments below 0.5\,K would be necessary to clarify this point.

\section{Discussion}

The A-site chemical substitution in the QSL candidate NaYbO$_2$ has two main implications. First, the $g$-tensor of Yb$^{3+}$ becomes symmetric. Second, magnetic interaction strength decreases. The latter effect can be ascribed to the increased Yb--O--Yb angle. Indeed, with the angle of $96.76^{\circ}$ KYbO$_2$ takes an intermediary position between NaYbO$_2$ ($95.72^{\circ}$) and YbMgGaO$_4$ ($100.36^{\circ}$). The absolute value of the powder-averaged Curie-Weiss temperature $\theta_{CW}$ correspondingly decreases from 6.4\,K in NaYbO$_2$~\cite{Ding2019} to 5.4\,K in KYbO$_2$ and 2.3\,K in YbMgGaO$_4$~\cite{Li2015PRL}. This reduction further manifests itself in the lower saturation field and in the shrinkage of the plateau region in the phase diagram (Fig.~\ref{fig:5}).

The effect on the $g$-value is more subtle. It is probably related to the change in the local environment of Yb$^{3+}$ that determines the CEF levels and the composition of the ground-state Kramers doublet. Interestingly, in KYbO$_2$ the trigonal distortion of the YbO$_6$ octahedra increases, but at the same time the alkaline-metal ions move further away from Yb$^{3+}$. The A--Yb distance increases from 3.36\,\r A in NaYbO$_2$ to 3.65\,\r A in KYbO$_2$. A similar effect was reported in AgYbO$_2$~\cite{Sichelschmidt2020} where the smaller Na$^+$ is also replaced by the larger ion (Ag$^+$). However, that sample showed a very broad ESR line with $g=3.1(5)$, while our KYbO$_2$ sample reveals a more narrow line, which is comparable in width with the line of NaYbO$_2$, yet isotropic with $g=3.08(3)$.

Isotropic nature of the local moment simplifies interpretation of the data collected on powder samples. Indeed, easy-plane anisotropy of the $g$-tensor for NaYbO$_2$ ($g_{\|}=1.75(3)$ and $g_{\perp}=3.28(8)$) suggests that the magnetization curve of this compound should be a superposition of the signals obtained for different directions of the applied field. This explains the residual curvature observed in $M(H)$ between 11 and 16\,T because the $H\|c$ component may have a higher saturation field due to the lower $g$-value. This anisotropy should also affect the magnetization value at the plateau and precludes its unambiguous assignment to the \textit{uud} ($M_{\rm sat}/3$) or \textit{uuud} ($M_{\rm sat}/2$) phases. In the case of KYbO$_2$, we do not observe any significant residual curvature above the saturation field of 9\,T, as the $g$-value is isotropic. Then the magnetization of about 0.7\,$\mu_B$/f.u. at the plateau serves as the fingerprint of the \textit{uuud} state that distinguishes KYbO$_2$ from the AYbSe$_2$ selenides with the field-induced \textit{uud} order. 

The \textit{uuud} state is indeed expected~\cite{Ye_2017O} in triangular antiferromagnets with $J_2/J_1>0.125$ where $J_1$ and $J_2$ stand for the nearest-neighbor and next-nearest-neighbor couplings, respectively. A sizable $J_2$ is thus a precondition for the formation of the field-induced \textit{uuud} state. Recent neutron scattering studies suggested $J_2/J_1=0.04-0.05$ in the AYbSe$_2$ selenides~\cite{Scheie2022,Xie2022}. This interaction regime is consistent with the $\frac13$ magnetization plateau and the field-induced \textit{uud} state observed in these compounds. On the other hand, it is plausible that the oxides should have a larger $J_2$ thanks to the shorter Yb--Yb distances. {Whereas the field-induced order in NaYbO$_2$ may be still of the \textit{uud} type, the lower value of $J_1$ in KYbO$_2$ should facilitate the $J_2/J_1>0.125$ regime and the $\frac12$ plateau typical of the \textit{uuud} order. Moreover, if the coupling $J_2$ is sensitive to the linearity of the Yb--O$\ldots$O--Yb pathway, as expected from the Cu$^{2+}$ compounds~\cite{prishchenko2017}, one should anticipate the increased $J_2$ in KYbO$_2$ with the Yb--O--O angle of $138.6^{\circ}$ compared to NaYbO$_2$ with its angle of $137.8^{\circ}$. This will further increase the $J_2/J_1$ ratio.}

{ The assignment of $J_2/J_1>0.125$ calls for a further theoretical study of this regime. Experimental magnetization curve may serve as a useful reference, as it shows a characteristic nonlinear behavior between zero field and the onset of the plateau [Fig.~\ref{fig:3}(b)], as opposed to the classical linear behavior. Another interesting aspect is the possible onset of magnetic order already at zero field that may be inferred from the divergence of the ESR linewidth (Fig.~\ref{fig:2}). The scale of this tentative transition is given by $T_c=0.15$\,K that, however, lies well below the temperature range covered in the present work.}

In summary, the A-site substitution in the QSL candidate NaYbO$_2$ leads to an anisotropic tensile strain and affects different aspects of the magnetism. Magnetic couplings are reduced following the increase in the Yb--O--Yb angles. This results in the lower saturation field and in the shift of the intermediate field-induced ordered state toward lower fields and temperatures. Additionally, the change in the local environment of Yb$^{3+}$ renders the $g$-tensor almost isotropic. A partial A-site substitution may then be a viable strategy for a fine tuning of magnetic interactions in triangular antiferromagnets. We have also shown that KYbO$_2$ develops the $\frac12$ magnetization plateau, which is indicative of the \textit{uuud} field-induced order and may signal the sizable second-neighbor coupling $J_2$ in this compound.

\begin{figure}[h]
\centering
\includegraphics[width=0.95\linewidth]{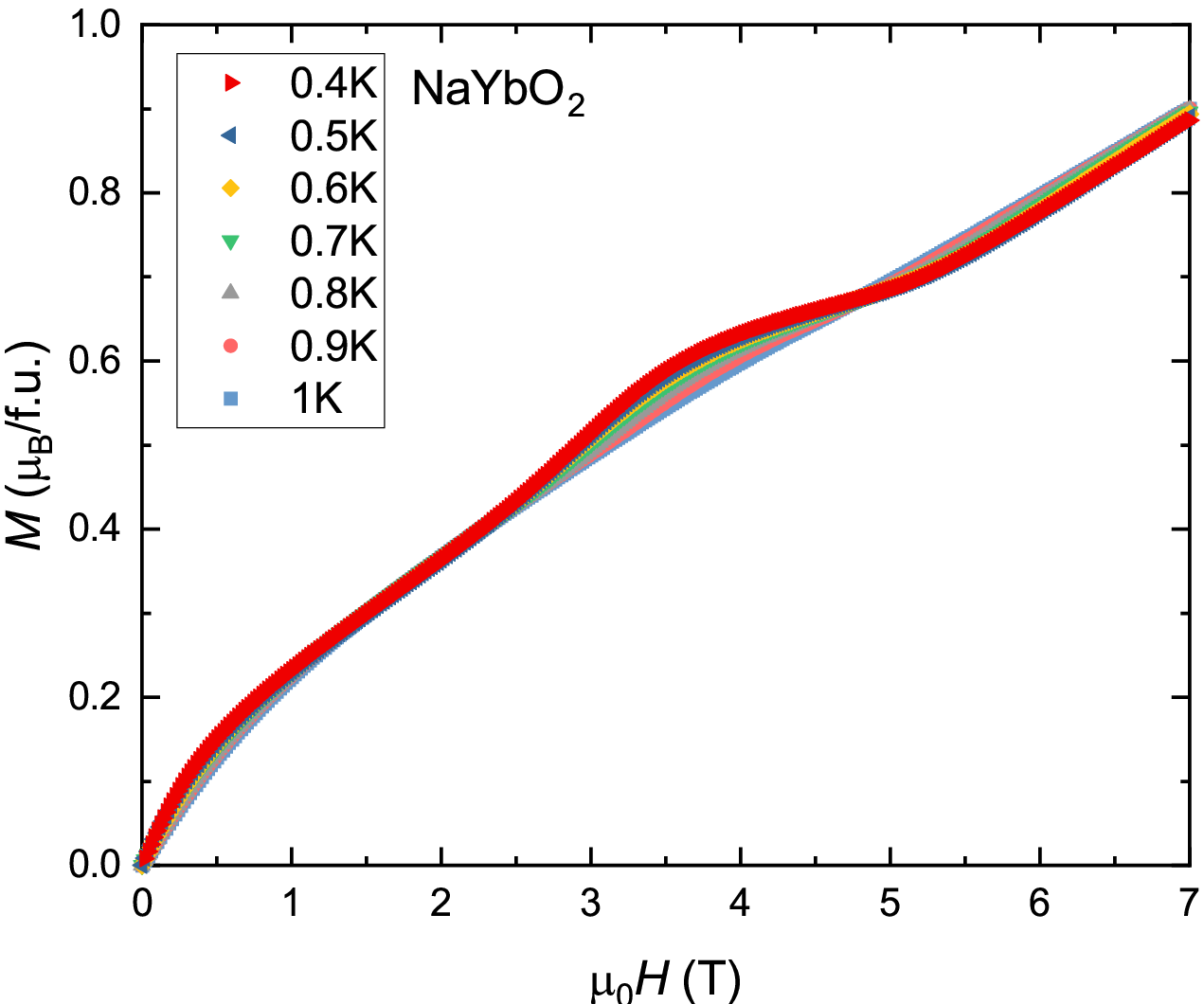}
\caption{Low-temperature magnetization of NaYbO$_2$. The plateau is located between 3 and 5.5\,T at about half of the saturation magnetization.}
  \label{fig:7}
\end{figure}

\acknowledgments
We thank Aleksandr Missiul, Francois Fauth, and Alexander Zubtsovskii for their help with the data collection at MSPD. This work was funded by the German Research Foundation (DFG) via the Project No. 107745057 (TRR80). We also acknowledge the support of the HLD at HZDR, member of European Magnetic Field Laboratory (EMFL).

\appendix*
\section{Additional data for NaYbO$_2$} \label{Appendix}
For a systematic comparison with KYbO$_2$, we performed additional measurements on NaYbO$_2$. These measurements extend the data that have been reported by us earlier~\cite{Ding2019}. Fig.~\ref{fig:6} shows specific heat of NaYbO$_2$ and confirms the absence of magnetic order in zero field. Applying magnetic field induces long-range order manifested by a $\lambda$-type anomaly above 3\,T. Similar to KYbO$_2$, the anomaly shifts to higher temperatures when increasing the field up to 5\,T, while for higher fields this trend is reversed. The position of this peak is used to track the phase boundary shown in Fig.~\ref{fig:5}. Additionally, the second peak  appears above 6\,T as a shoulder on the lower-temperature side, similar to the data reported in Ref.~\cite{Ranjith2019_NaYbO2}. We interpret this second peak as the transition induced by the field $H\|c$ because the lower component of the $g$-tensor weakens the effect of the out-of-plane field and, correspondingly, a higher field strength would be needed to stabilize field-induced order. Since $g_{\perp}=3.28(8)$ of NaYbO$_2$ is similar to the isotropic $g=3.08(3)$ of KYbO$_2$, while $g_{\|}=1.75(3)$ is much lower, we used the first peak when tracing the phase boundary. We also measured the low-temperature $M(H)$ curves for NaYbO$_2$ (Fig.~\ref{fig:7}) and {traced the field range of the plateau using maxima and minima in the second derivative.}

\bibliography{KYbO2_manuscript}

\end{document}